\documentclass[mathphys, Numbered]{jnl}
\usepackage{graphicx}%
\usepackage{multirow}%
\usepackage{amsmath,amssymb,amsfonts}%
\usepackage{amsthm}%
\usepackage{mathrsfs}%
\usepackage[title]{appendix}%
\usepackage{xcolor}%
\usepackage{textcomp}%
\usepackage{manyfoot}%
\usepackage{booktabs}%
\usepackage{algorithm}%
\usepackage{algorithmicx}%
\usepackage{algpseudocode}%
\usepackage{listings}%
\algnewcommand\algorithmicto{\textbf{to}}
\algnewcommand\RETURN{\State \textbf{return} }

\begin{document}

\title[Article Title]{Image-guided Computational Holographic Wavefront Shaping}

\author[1]{\fnm{Omri} \sur{Haim}}
\equalcont{These authors contributed equally to this work.}

\author[1]{\fnm{Jeremy} \sur{Boger-Lombard}}
\equalcont{These authors contributed equally to this work.}

\author[1]{\fnm{Ori} \sur{Katz}}\email{orik@mail.huji.ac.il}

\affil[1]{\orgdiv{Institute of Applied Physics}, \orgname{The Hebrew University of Jerusalem}, \orgaddress{\city{Jerusalem}, \postcode{9190401}, \country{Israel}}}


\abstract{ Optical imaging through scattering media is an important challenge in a variety of fields ranging from microscopy to autonomous vehicles. 
While advanced wavefront shaping techniques have offered significant breakthroughs in the past decade, current techniques still require a known guide-star and a high-resolution spatial-light-modulator (SLM), or a very large number of measurements, and are limited in their correction field-of-view. 
Here, we introduce a guide-star free noninvasive approach that is able to correct more than $3\cdot 10^5$ scattered modes using just $100$ holographically measured scattered random light fields. 
This is achieved by computationally emulating an image-guided wavefront-shaping experiment, where several 'virtual SLMs' are simultaneously optimized to maximize the reconstructed image quality. 
Our method shifts the burden from the physical hardware to a digital, naturally-parallelizable computation, leveraging  state-of-the-art automatic-differentiation optimization tools used for the training of neural-networks. 
We demonstrate the flexibility and generality of this framework by applying it to imaging through various complex samples and imaging modalities, including anisoplanatic multi-conjugate correction of highly scattering layers, lensless-endoscopy in multicore fibers, and acousto-optic tomography.
The versatility, effectiveness, and generality of the presented approach have great potential for rapid noninvasive imaging in diverse applications.}

\maketitle

\section*{Introduction}\label{intro_sec}

High resolution optical imaging and light control in complex media are crucial for a diverse array of applications, spanning from microscopy \cite{ntziachristos2010going, bertolotti2022imaging}, and astronomy \cite{davies2012adaptive} to telecommunications \cite{9829750}. 
However, the performance of even the most advanced optical systems is still limited by the inherent random scattering of light in complex samples. Examples include scattering in biological tissues, which limits the penetration depth of optical microscopes to less than a millimeter, and scattering in dense fog, which limits the viewing range of even the most advanced laser-based systems.
Great research efforts to overcome scattering over the years have resulted in a variety of techniques such as confocal-microscopy, multi-photon microscopy, optical coherence tomography (OCT), and acoustically-mediated tomographies \cite{ntziachristos2010going, bertolotti2022imaging}. While these techniques found widespread use, they are all still limited in their penetration-depth and/or resolution since they do not attempt to correct scattering but rather avoid it by either detecting only un-scattered ‘ballistic’ light or by relying on ultrasound to provide an imaging resolution orders of magnitude inferior to the optical one \cite{ntziachristos2010going, bertolotti2022imaging}.

While scattering of coherent light is a complex and random process, it is also a deterministic one, and it can be physically undone by high resolution wavefront-shaping using computer controlled spatial light modulators (SLMs) or nonlinear phase-conjugating crystals \cite{vellekoop2007focusing, mosk2012controlling, horstmeyer2015guidestar, cao2022high, bertolotti2022imaging}.  
In the past decade, 'wavefront shaping' has proved effective in physically correcting multiple scattering even when the number of SLM controlled pixels (degrees of freedom, DoF) is orders of magnitude smaller than the number of scattered modes, as is the case in e.g. deep tissue imaging. Strikingly, the price of the imperfect correction is only in the image contrast and signal level, and not in resolution, which remains diffraction-limited  \cite{vellekoop2007focusing, bertolotti2022imaging,mosk2012controlling}, a qualitative difference from adaptive-optics \cite{booth2012adaptive, li2015conjugate}.

Although the performance of wavefront-shaping in laboratory experiments is well beyond the capabilities of any conventional microscope \cite{park2015high, papadopoulos2017scattering, badon2019distortion,kang2023tracing, may2021fast}, a fundamental gap exists in wavefront-shaping ability to address the requirements of speed, field of view (FoV), and object complexity of many practical imaging applications, restricting wavefront-shaping from realizing its potential impact and widespread use. 
The two main reasons for this gap are the requirements for the presence of  known ‘guide-stars’ at the target plane that are used to determine the wavefront corrections \cite{horstmeyer2015guidestar, bertolotti2022imaging}, or the necessity to perform a very large number of controlled wavefront measurements to acquire the reflection-matrix of the medium \cite{kang2023tracing, badon2019distortion}. 

When sufficient isoplanatism is present, i.e. when the scattering is tilt-invariant to a sufficient degree, also known as the 'optical memory-effect' \cite{freund1988memory, bertolotti2022imaging}, decomposition of the reflection matrix \cite{badon2019distortion, kang2023tracing}, or correction of the  guide-star wavefront distortion, allow imaging over the memory-effect range through highly scattering samples, such as biological tissues \cite{badon2019distortion, kang2023tracing, horstmeyer2015guidestar, bertolotti2022imaging} and multicore fibers \cite{choi2022flexible}. 
However, in practical scenarios it is many times impossible or extremely challenging to implant guide-stars or to acquire a sufficiently large reflection-matrix within the decorrelation time of naturally dynamic samples such as live specimens \cite{cheng2023high}, rendering the application of the state-of-the-art wavefront-shaping techniques challenging.  
Single-shot computational approaches that utilize the memory effect for noninvasive speckle-correlations imaging \cite{bertolotti2012non, katz2014non, porat2016widefield} have been developed to circumvent the need for guidestars or multiple measurements, but they face difficulties in reconstructing complex objects, or objects that extend beyond the memory-effect \cite{rosenfeld2021acousto}.

Recently, to overcome the requirement for guide-stars, noninvasive focusing and imaging was performed via fluorescence variance optimization\cite{daniel2019light,boniface2019noninvasive}
image-guided optimization \cite{yeminy2021guidestar}, or non-negative matrix factorization \cite{boniface2020non}. These approaches remove the need for guidestars entirely, however, they still rely on a large number of measurements and a physical correction device, resulting in a very slow correction speed, as well as isoplanatism-limited FoV.
Here, we introduce a holography-based computational technique that enables noninvasive high-resolution imaging through complex media without the use of guide stars or SLMs, and without the knowledge or measurement of the medium reflection-matrix.  
Specifically, using only $<200$ holographic measurements, we determine the $>350,000$ DoF optimal wavefront correction that maximizes a computationally-reconstructed image quality. The optimization is performed in a highly-parallel fashion using an automatic-differentiation based gradient-descent algorithm commonly used for the training of deep neural networks.  

The proposed approach builds upon image-guided wavefront shaping \cite{yeminy2021guidestar}, where the target object serves as its own 'guide-star', but here we replace the slow and limited-DoF physical wavefront-shaping hardware by a computational 'virtual SLM', and the sequential wavefront optimization process by a naturally-parallelizable and highly efficient computational optimization. 
In addition to shifting the burden from the physical hardware to a digital parallel computational process, the approach does not require careful control or knowledge of the illumination or labeling of the object. It also enables anispolanatic correction of thick samples by 'placing' multiple virtual SLMs at different conjugate planes,  in an analogous manner to multi-conjugate adaptive optics \cite{davies2012adaptive,li2015conjugate}.

Interestingly,  computational image-guided correction is also employed in computational adaptive-optics (CAO) \cite{tippie2011high, adie2012computational}, and in digital autofocus in Synthetic Aperture Radar \cite{schulz2006optimal}. 
However, its adaptation to wavefront-shaping of multiple-scattering is fundamentally different due to the complexity of the distortions and DoFs available for correction. This is analogous to the difference between wavefront-shaping and adaptive-optics: CAO is used to correct low-order aberrations usually described by a few tens or hundreds of parameters, while here we tackle diffusive scattering to more than $10^6$ modes, and apply corrections utilizing up to $360,000$ controlled DoF. Most importantly for image-guided optimization, while the conventional starting point in CAO (and conventional adaptive optics) is a somewhat blurred image \cite{debarre2007image, tippie2011high, adie2012computational}, in computational wavefront shaping it is a highly complex, and potentially very low contrast, speckle pattern that has no resemblance to the target object.

To showcase the effectiveness and flexibility of our approach we experimentally apply it to various complex media and measurement approaches, including highly scattering layers, flexible multi-core fibers, and acousto-optical tomography datasets. The simple gradient-descent code naturally lends itself to adaptation to different targets or imaging modalities.

Over 70 years ago, Dennis Gabor invented holography as an approach to overcome aberrations in electron microscopes \cite{gabor1949microscopy}. Overcoming scattering using computation can be thought of as the digital-age analogue of Dennis Gabor’s vision.


\section*{Results}\label{results_sec}
\subsection*{Principle}\label{fig:principle}
\begin{figure*}[h!]
    \centering
    \includegraphics[width=0.97\textwidth]{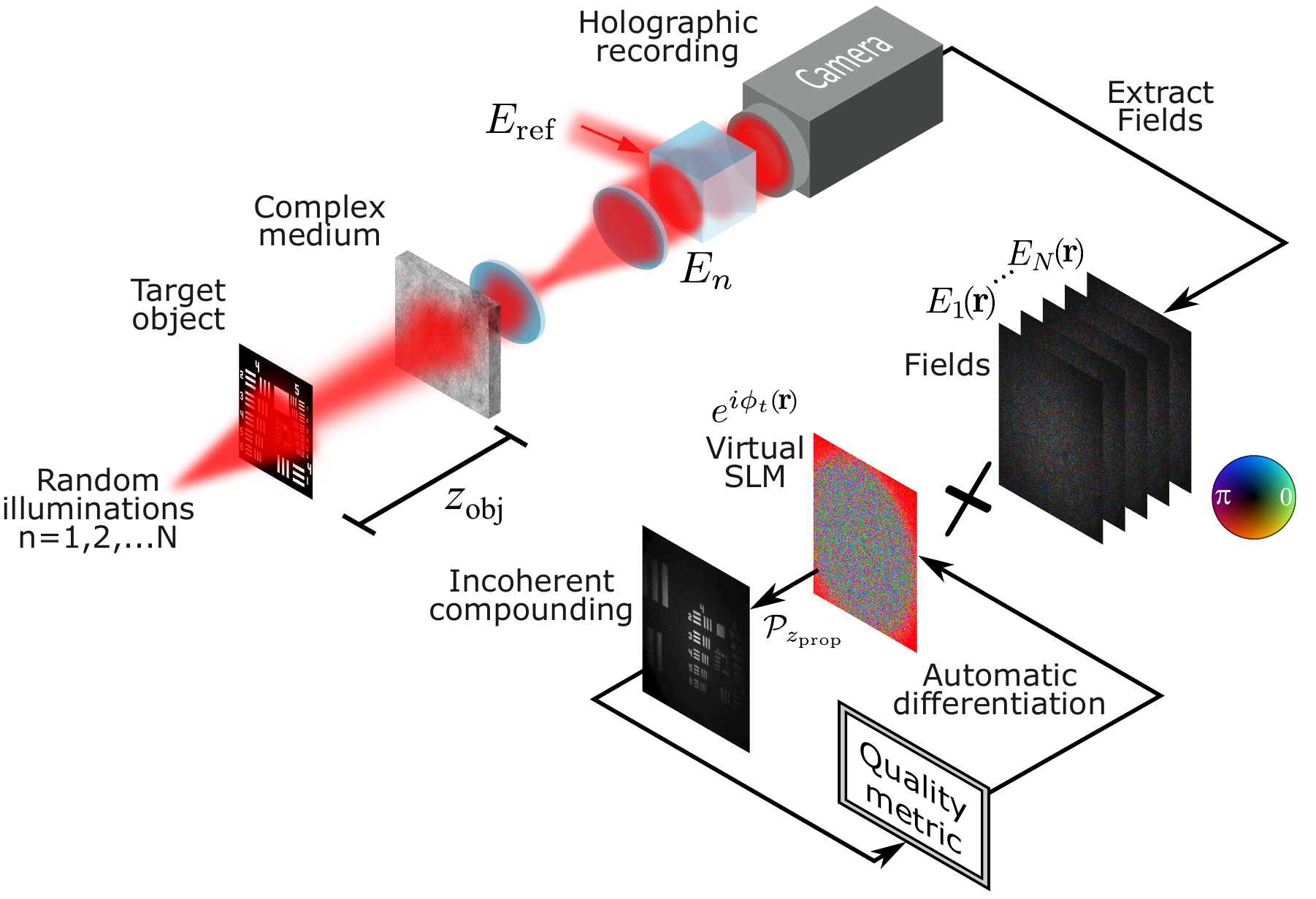}
    \caption{\textbf{Image-guided computational wavefront shaping}. 
    The concept is divided into two parts: an acquisition sequence (top) where $N<200$ scattered light fields that have diffused from the target object through the complex medium are holographically recorded by a high-resolution camera, and a computational reconstruction step (bottom) where the holographically-recorded fields are computationally traced-back to the estimated object plane, using a correction phase mask $\phi_t(\mathbf{r})$, forming a 'virtual SLM' and computational back-propagation.
    The propagated fields' intensities are then incoherently compounded and an image quality metric is calculated. 
    Using an automatic-differentiation gradient-descent optimization algorithm, commonly used for training of deep neural networks, the correction phase-mask, $\phi_t(\mathbf{r})$, is optimized to maximize the image quality metric. The end result is both the complex correction phase mask and the corrected object image.}
\end{figure*}

The principle of image-guided computational wavefront-shaping together with a schematic representation of the experimental setup and a sample experimental result are presented in Fig.\ref{fig:principle}.  A hidden object, placed behind or inside a highly scattering complex medium, is illuminated by random unknown speckle patterns.
A high-resolution camera that is placed outside the complex medium holographically records the scattered light field that has diffused through the complex medium. The holographic recording is repeated for each of a few tens to 200 of random unknown spatially-coherent speckle illuminations. Each hologram of the scattered light field at the proximal facet of the scattering medium is captured in a single shot via off-axis holography (see Supplementary Fig.1).
Following the holographic acquisition the object image is reconstructed by a computational physical-model based optimization process inspired by image-guided wavefront-shaping  \cite{yeminy2021guidestar} (Fig.\ref{fig:principle}). 

In image-guided wavefront-shaping a physical wavefront correction is found by varying an SLM phase-pattern to optimize the spatially-incoherent image quality metric \cite{yeminy2021guidestar}. 
Here, we find the computational wavefront correction that digitally reconstructs the image of the unknown target with the highest quality (here image variance, see Methods). Given the $N$ holographically-measured proximal fields, $E_{n}(\mathbf{r}) $, where $n=1...N$, the iterative computational inversion process aims at simultaneously finding both the correction phase mask of a virtual SLM conjugated to the complex medium surface, and the object image (Fig.\ref{fig:principle}).
This is performed by computationally emulating an image-guided wavefront-shaping experiment, where the spatial phases of a 'virtual SLM' correction are optimized to maximize the quality of the reconstructed incoherently-compounded object image (Fig.\ref{fig:principle}). 

The object image is reconstructed by digitally reversing the physical propagation through the complex medium, most simply modeled as a thin phase mask (see Methods), a simplification that will be relaxed below. Specifically, at each iteration of the algorithm, $t$, the incoherently-compounded object image is reconstructed by tracing back each of the measured fields through a correction mask,  $e^{-i\phi_{t}(\textbf{r})}$, and digital back-propagation of a distance $-z_\mathrm{prop}$ to the estimated plane where the object lies. 
Since the computationally reconstructed image quality depends in a well defined fashion on the virtual SLM phase pattern (see Methods), efficient automatic-differentiation based gradient-descent search after the phase pattern yielding the optimal image quality can be performed. This is done in a straightforward fashion using a GPU-optimized Python library developed for the training of deep neural networks \cite{paszke2017automatic}. 

Importantly, while the digital reconstruction process is based on back-propagating the recorded fields by the estimated distance to the object plane, knowledge of this distance is not required since the phase-mask optimization  also effectively finds the object position through the possible addition of a spherical phase mask in a computational autofocus process.

Applying a digital correction with a 'virtual SLM' rather than a physical SLM carries several important advantages as we demonstrate below: it allows for a fast acquisition process that is adapted to dynamically varying media, 
it allows parallel optimization of a very large number of controlled DoFs, orders of magnitude more efficient than current wavefront-shaping optimization approaches such as genetic algorithms. 
Finally, the computational correction allows the use of multiple virtual SLMs  conjugated to different planes in the complex medium. This enables the correction of anisoplanatic distortions and imaging objects that extend beyond the memory-effect range. All of these features are not possible with physical iterative wavefront optimization and carry great potential for many applications.

\subsection*{Experimental imaging through highly scattering layers}\label{experimental_diffuser_imaging}

\begin{figure*}[h!]
    \centering
    \includegraphics[width=\textwidth]{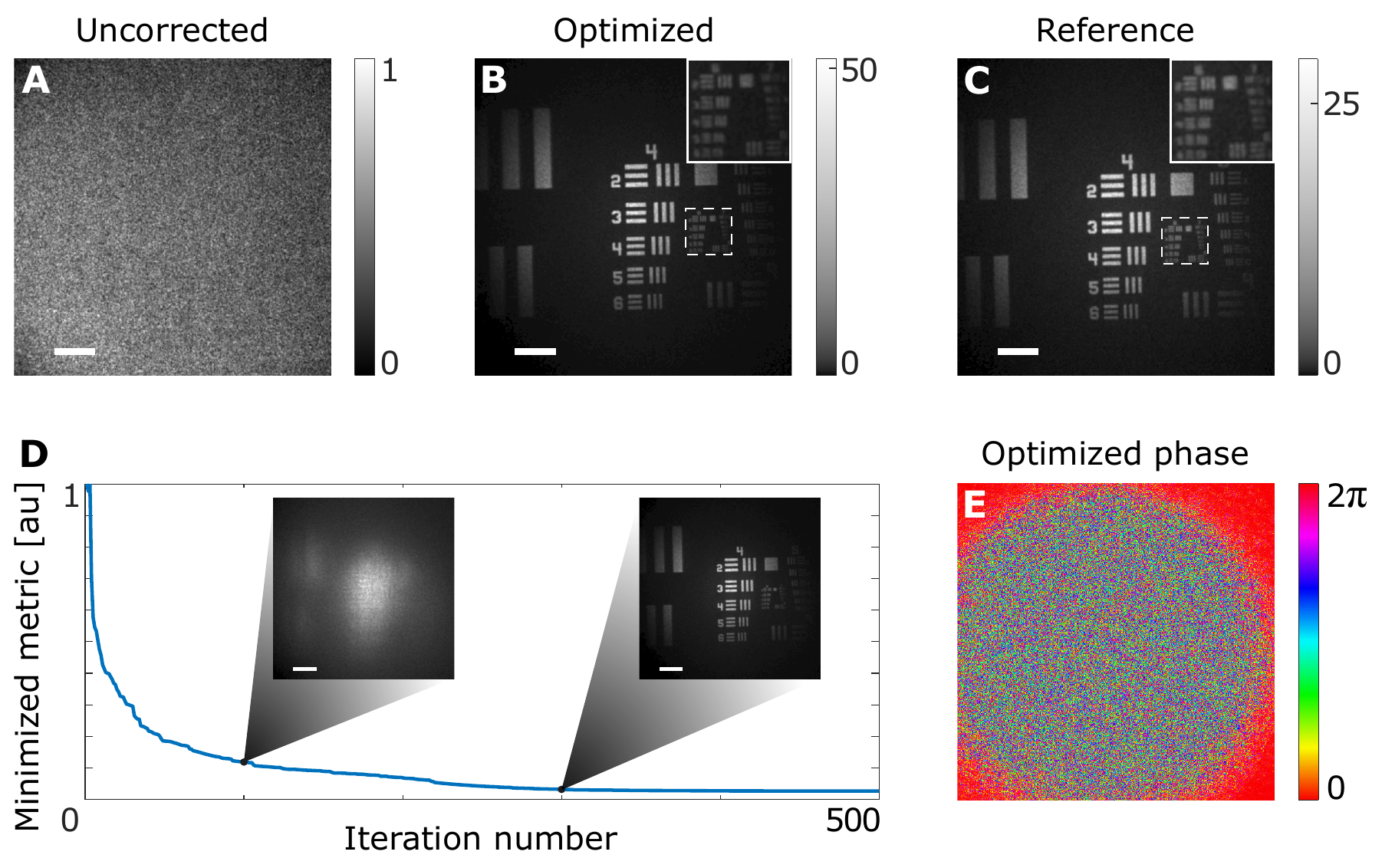}
    \caption{\textbf{Experimental imaging through a highly scattering diffuser using $600\times600$ degrees of freedom}.  (\textbf{A}) Object-plane reconstruction using the uncorrected fields. (\textbf{B}) Optimized object-plane reconstruction. (\textbf{C}) A reference corrected reconstruction using an invasively measured reference phase of the complex medium taken without the target. (\textbf{D}) Evolution of the minimized image-metric, maximizing the image variance (see Methods) as a function of the optimization iterations, with intermediate reconstructions at the $100^\mathrm{th}$ and $300^\mathrm{th}$ iterations. (\textbf{E}) The final optimized correction phase mask, $\phi_t(\mathbf{r})$,  yielding the optimized reconstruction in (B). Scale bars, 300$\mu$m.}
    \label{fig:optim_evolution}
\end{figure*}

As a first demonstration of guidestar-free, high DoF widefield computational wavefront-shaping we present results of imaging a USAF resolution target through a highly scattering diffuser ($60^{\circ}$ light shaping diffuser, Newport). 
Using only 100 holographic measurements we are able to effectively correct the wavefront distortion using a virtual SLM with $600\times600$ DoFs. 
The results presented in Fig.\ref{fig:optim_evolution} show that while the uncorrected reconstructed image is a low contrast diffusive blur (Fig.\ref{fig:optim_evolution}A), the computationally corrected image is a high-contrast diffraction-limited image (Fig.\ref{fig:optim_evolution}B) that reveals all the details of the target. This is verified by comparison to the reconstructed image of the target when the wavefront correction is recorded in an invasive reference measurement prior to placing the target (Fig.\ref{fig:optim_evolution}C) \cite{li2019rapid}. 

As mentioned above, it is not required to know in advance the object distance, $z_\mathrm{obj}$. The optimization successfully recovers the object even when a different back-propagation distance, $z_\mathrm{prop}$,  is used. It compensates for the propagation difference by an additional spherical phase mask correction (see Supplementary Fig.2), 
in a manner analogous to digital autofocus \cite{schulz2006optimal}. 

Fig.\ref{fig:optim_evolution}D presents the evolution of the image quality metric during the iterations together with two intermediate reconstructed images (Fig.\ref{fig:optim_evolution}D, insets), and Fig.\ref{fig:optim_evolution}E presents the optimized phase mask containing more than 300,000 DoFs. 

The results of Fig.\ref{fig:optim_evolution} may raise the question: how can such a large number of parameters ($N_\mathrm{DoF}=3.6\cdot 10^5$) be determined from a small number of captured frames ($N=100$)? The answer is that each of the $N=100$  holograms is a megapixel frame composed of $N_\mathrm{pixels}\gg10^5 $ complex-valued field measurements. Thus, the number of total measurements is orders of magnitude larger than the number of optimized DoFs: $N\cdot N_\mathrm{pixels}\gg N_\mathrm{DoF}$.

\begin{figure*}[h!]
    \centering
    \includegraphics[width=0.97\textwidth]{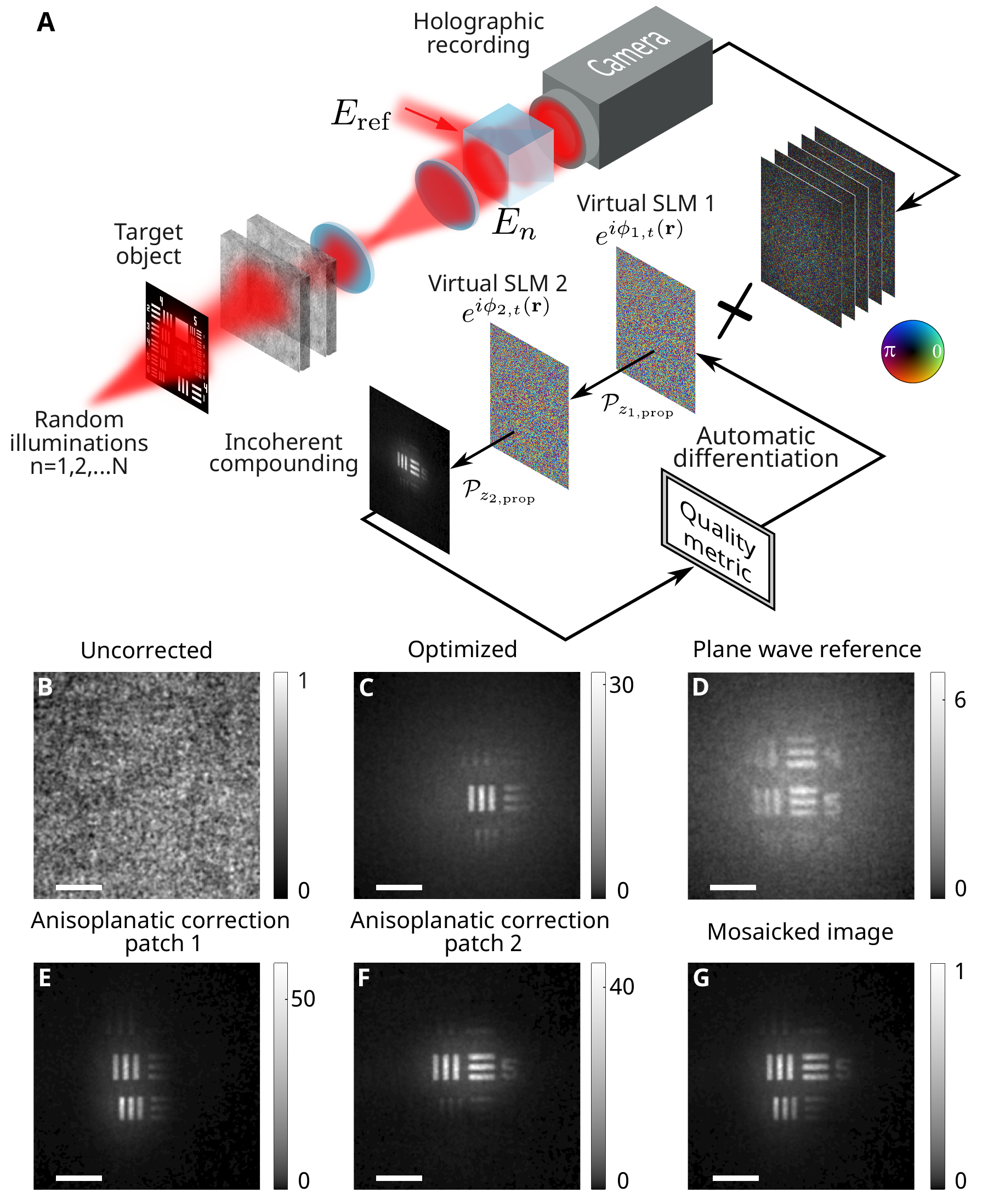}
    \caption{\textbf{Multi-conjugate anisoplanatic correction through a volumetric scattering sample}. (\textbf{A}) Schematic of the experiment where the complex sample is composed of two separated scattering diffusers, having a limited memory-effect. (\textbf{B}) Uncorrected object plane reconstruction. (\textbf{C}) Optimized object plane reconstruction using a single virtual SLM phase mask allows the correction of a single isoplanatic patch. (\textbf{D}) Reference reconstruction using a correction based on an invasively measured wavefront of the complex medium illuminated by a plane wave. (\textbf{E}) Optimized multi-conjugate object-plane reconstruction using two correction phase masks, allowing anisoplanatic correction of a wider FoV. (\textbf{F}) Same as (E) when optimizing the metric on a different region of interest. (\textbf{G}) Mosaicked image of the results in (E) and (F).
    Scale bars 300$\mu$m.}
    \label{fig:2 diffusers}
\end{figure*}

As in conventional wavefront-shaping, widefield imaging with a single SLM correction implicitly assumes isoplanatic disortions, i.e that the target object does not extend beyond the memory-effect range \cite{bertolotti2022imaging}. 
Going beyond isoplanatic scattering requires the use of more than a single SLM, since a wavefront correction performed by such a two-dimensional planar device cannot undo the scattering of a volumetric three-dimensional medium. This limitation is tackled in astronomical observations by multi-conjugate adaptive-optics (MCAO), where multiple deformable-mirrors are conjugated to different aberrating layers of the atmosphere \cite{beckers1988increasing}. While realizing multi-conjugate wavefront-shaping is experimentally challenging it is straightforward to perform in our computational approach, as we demonstrate in Figure \ref{fig:2 diffusers}.

Figure \ref{fig:2 diffusers} presents experimental results of imaging through a complex thick sample that introduces anisoplanatic scattering, i.e. a limited memory-effect. 
The experimental setup remains the same, but the complex sample now consists of two separated optical diffusers (Fig.\ref{fig:2 diffusers}A). 
Performing the computational correction with a single virtual SLM as before  still retrieves a small part of the hidden target that is contained within the memory-effect range (Fig.\ref{fig:2 diffusers}C), similar to the results of conventional wavefront-shaping \cite{yeminy2021guidestar}. Nonetheless, the computational image guided correction can be easily extended to multiple virtual SLMs conjugated to different planes in the thick scattering sample (Fig.\ref{fig:2 diffusers}A, see Supplementary section 7).
The reconstruction results with multi-conjugate computational wavefront shaping, using two correcting phase masks conjugated to the estimated planes of the two diffusers are presented in Fig.\ref{fig:2 diffusers}E-G. As expected, the corrected FoV is extended beyond that of the single plane correction (Fig.\ref{fig:2 diffusers}C). Moreover, the correction of different regions is possible by using different regions of the image for the metric calculation  (Fig.\ref{fig:2 diffusers}E, F). Mosaicking these results enables imaging of an even larger FoV (Fig.\ref{fig:2 diffusers}G).  
Such multi-conjugate correction of severe wavefront distortions has not been demonstrated with physical SLMs, as it is technically challenging, and has only very recently demonstrated computationally using a reflection-matrix based approach \cite{kang2023tracing}.

\newpage
\subsection*{Application in lensless endoscopy}\label{experimental_mcf_imaging}

\begin{figure*}[h!]
    \centering
    \includegraphics[width=0.97\textwidth]{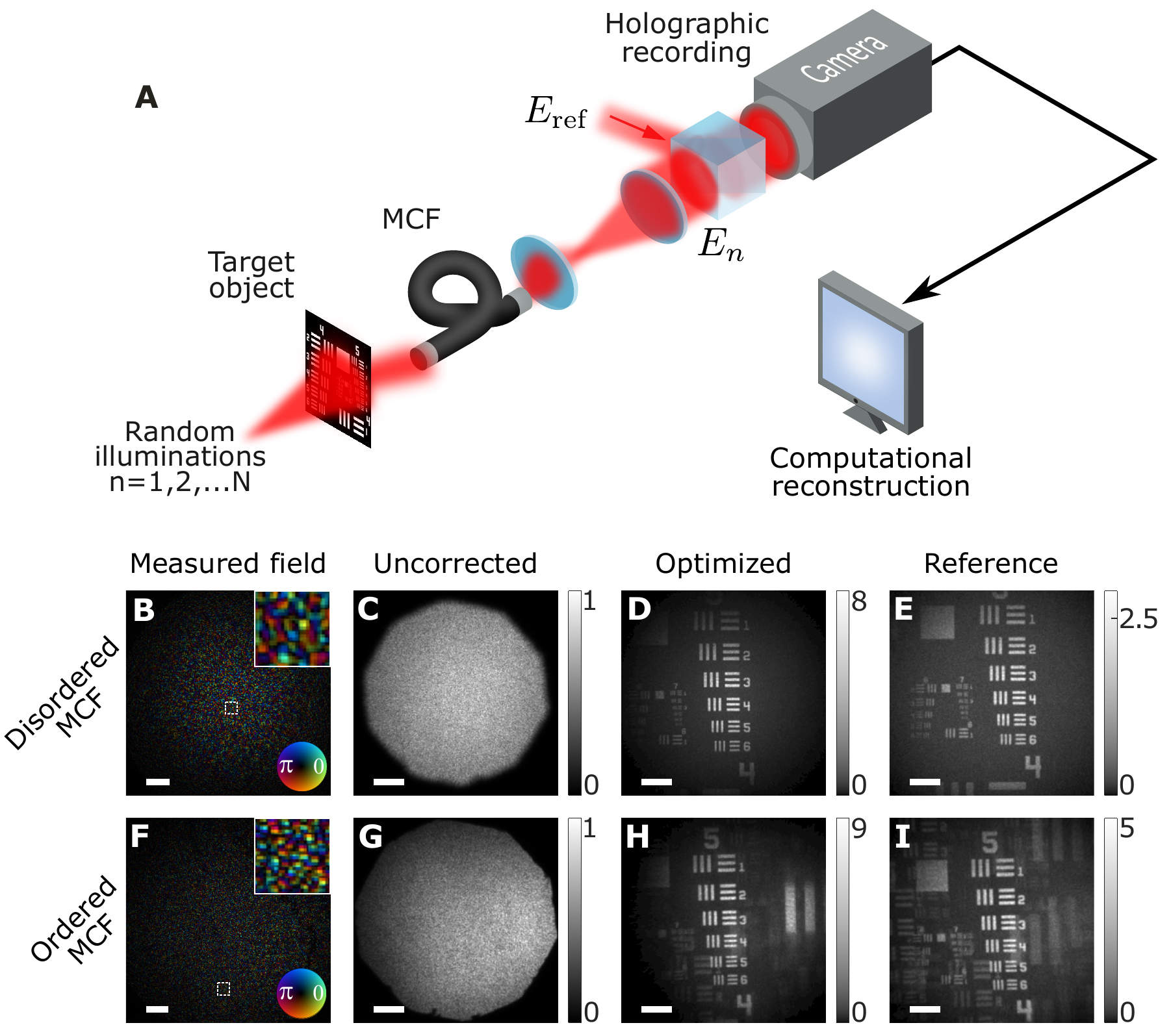}
    \caption{\textbf{Computational image-guided lensless endoscopy through multi-core fibers} (MCF). \textbf{(A)} Experimental setup: A target object situated at various small distances (here $z_\textrm{obj}= 5$mm) from the distal facet an MCF is illuminated by varying unknown random speckle patterns. A camera holographically records the distorted fields at the proximal facet of the fiber over 100 illumination realizations. The recorded fields are fed as input to the computational reconstruction algorithm along with the estimated distance of the object (\ref{fig:principle}B). \textbf{(B-E)} Results using a commercial MCF having disordered and inhomogeneous cores. (B) A single recorded proximal field. (C) Reconstructed object-plane before correction. (D) Optimized object-plane reconstruction. (E) Reference reconstruction using an invasively measured phase distortion of the MCF without a target present. \textbf{(F-I)} Same as (B-E) using a commercial MCF with ordered homogeneous cores, displaying reconstructed replicas due to the periodic spatial sampling of the MCF cores. Scale bars 200$\mu$m.}
    \label{fig:fiber_results}
\end{figure*}

The computational image-guided framework is versatile and can be applied to a variety of different image modalities and distortion mechanisms. In this section, we demonstrate this by applying our approach to lensless endoscopic imaging. 
Lensless endoscopes offer a promising solution for minimally invasive microendoscopy due to their smaller footprint and dynamic 3D imaging capabilities \cite{andresen2013two,ploschner2015seeing}. One potential platform for lensless endoscopy is multi-core fiber (MCF) bundles, consisting of up to thousands of individual cores \cite{andresen2013two,weiss2018two}. Traditionally, the fiber distal tip is placed adjacent to the imaged object or conjugated to it using a distal lens, resulting in a fixed focal plane and a larger footprint. To achieve 3D imaging of objects located away from the fiber distal facet, without a distal lens, the random spatial phases of different cores need to be corrected. This has been accomplished by wavefront shaping either through invasive access to the distal side \cite{andresen2013two}, using nonlinear feedback signals \cite{weiss2018two}, or image-guided wavefront shaping \cite{yeminy2021guidestar}, all of which are either invasive or extremely slow.

To demonstrate the versatility of our approach we have applied it to several  lensless endoscopic imaging experiments performed with MCFs having either ordered and disordered array of cores, with negligible and non-negligible cross-talk, correspondingly. Importantly, this was achieved using the same setup and reconstruction algorithm as in Fig.\ref{fig:optim_evolution}. 
The only modification was the introduction of two different commercial MCFs instead of the scattering medium Fig.\ref{fig:fiber_results}A. 

Using $N=100$ captured holograms of the field at the proximal MCF facet (Fig.\ref{fig:fiber_results}B,F), our computational image-guided correction was able to correct the complex distortions (resulting in a low contrast diffusive blur Fig.\ref{fig:fiber_results}C,G ) of 50k and 18k MCF cores, respectively (Fig.\ref{fig:fiber_results}D,H). 
As expected, the reconstructed images through the MCF having a periodic ordered arrangement of cores (Schott - 1533385), contain replicas of the imaged object due to the periodic spatial sampling of the MCF cores (Fig.\ref{fig:fiber_results}D), which is not present in the disordered MCF bundle (Fujikura - FIGH-50-1100N) results (Fig.\ref{fig:fiber_results}H). Interestingly, these natural 'spatial-aliasing' replicas do not limit the convergence of the reconstruction algorithm, which was not adapted to their presence. Moreover, the computationally-optimized result has a higher contrast than the reconstruction using an invasively-measured reference correction, taken prior to presenting the target.

\subsection*{Application in acousto-optic tomography}\label{experimental_ao_imaging}
\begin{figure*}[h!]
    \centering
    \includegraphics[width=0.97\textwidth]{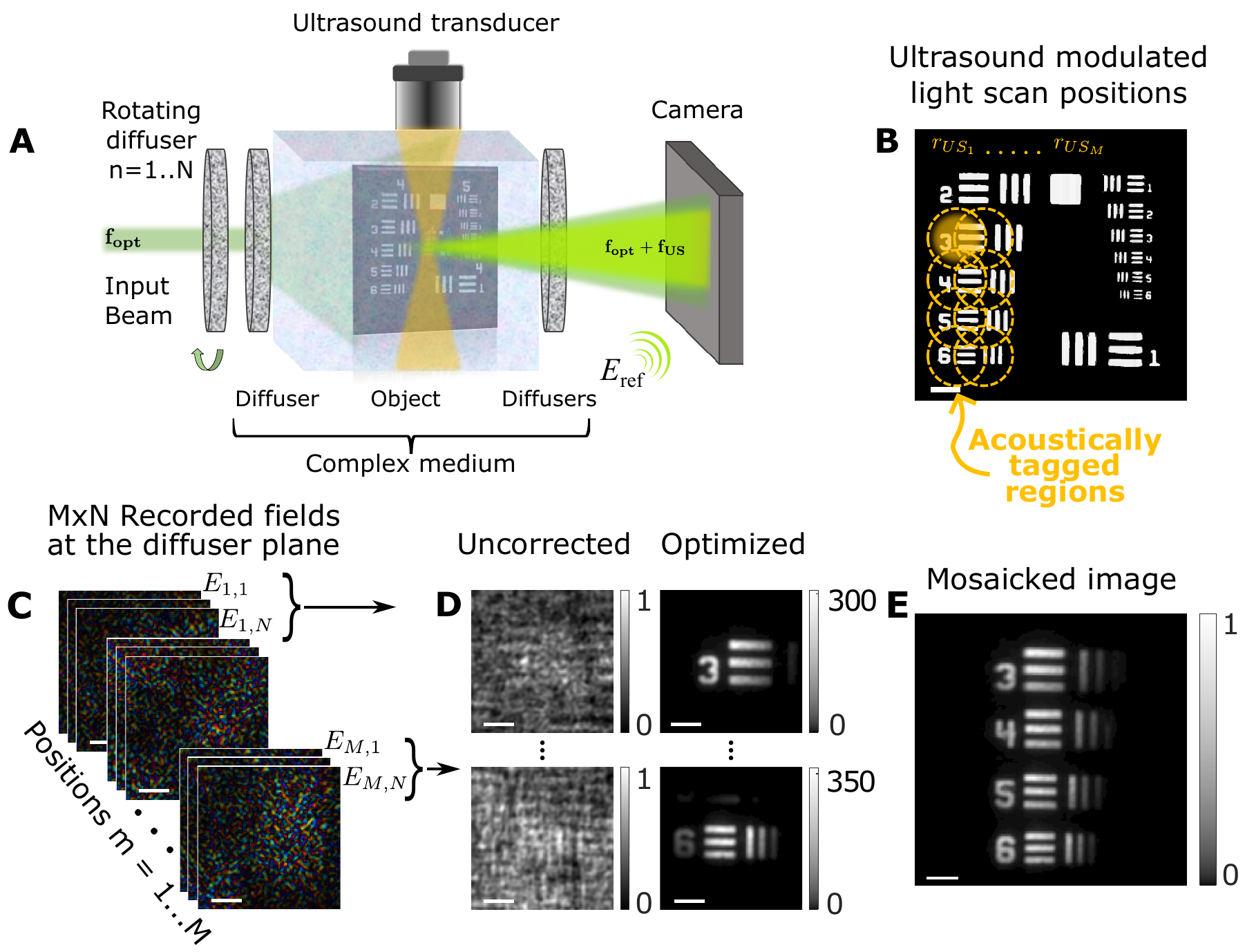}
    \caption{\textbf{Image guided computational wavefront shaping of ultrasonically-encoded light fields}. (\textbf{A}) Measurement setup: an acousto-optical imaging setup is equipped with a rotating diffuser for producing varying random illuminations. 
    An ultrasound pulse (yellow) is focused on several small sub-volumes of an extended object that is placed inside a complex sample.  The scattered ultrasonically-tagged light fields (light green) that originate from the selected sub-volumes are holographically recorded by a high resolution camera placed outside the sample.
    (\textbf{B}) The target object. The ultrasound beam focus positions are marked by dashed yellow circles. The object extends beyond the memory effect (isoplanatic) range, but the ultrasound focus size is smaller than the isoplanatic patch. 
    (\textbf{C}) The experimentally recorded scattered light fields. For each ultrasound focus position, $m=1..M$, $N=150$ scattered fields are captured, each one for a different random unknown illumination. 
    (\textbf{D}) Reconstruction of the object plane with uncorrected phase and with optimized phase correction at two acoustically tagged positions ($m=1,8$). 
    (\textbf{E}) Mosaicked image of all $M=8$ tagged regions reconstructions. Scale bars 100$\mu$m.}
    \label{fig:acousto_optics}
\end{figure*}

As a final demonstration of the applicability of the presented approach to a wide range of experimental settings and imaging modalities, we apply it to a holographic dataset from an acousto-optic tomography (AOT) experiment (Fig.\ref{fig:acousto_optics}).  
AOT is a state-of-the-art noninvasive deep-tissue optical imaging approach that combines the advantages of optical contrast with the near scatter-free propagation of ultrasound in soft tissues \cite{elson2011ultrasound,ntziachristos2010going}. 

In AOT a focused ultrasound spot is used to locally modulate light at chosen positions inside the sample (yellow circles in Fig.\ref{fig:acousto_optics}B). The ultrasound spot is generated and scanned inside the sample by an external ultrasonic transducer (Fig.\ref{fig:acousto_optics}A). The modulated, frequency-shifted light is detected outside the sample using, e.g. interferometry-based approaches \cite{elson2011ultrasound}. This enables the measurement of the total optical power traversing through the localized acoustic focus, and thus to imaging with a resolution given by the dimensions of the ultrasound focus. AOT is noninvasive and has a deep penetration but since the ultrasound focus dimensions are dictated by acoustic diffraction, the imaging resolution is several orders of magnitude inferior than the optical diffraction limit.  

Here, we utilize an experimental holographic dataset obtained from an AOT setup to reconstruct the image of an extended object hidden inside a complex sample with optical diffraction-limited resolution. 
This order-of-magnitude improvement in resolution over the state-of-the-art wavefront-shaping AOT techniques \cite{judkewitz2013speckle,katz2019controlling}, is made possible by exploiting the memory-effect that is present in the probed sample. 

The experimental setup is a conventional AOT setup that incorporates a rotating diffuser in the illumination arm. It is schematically presented in Fig.\ref{fig:acousto_optics}A, and described in detail in \cite{rosenfeld2021acousto}.  For each of  the $M=8$ ultrasound-probed regions, the ultrasound-modulated light fields are recorded by phase-shifting off-axis holography for $N=150$ random unknown illuminations  \cite{rosenfeld2021acousto}  (Fig.\ref{fig:acousto_optics}C).

Imaging the target object by wavefront-shaping without the acousto-optic modulation is extremely challenging for two reasons: first, the diffused illumination illuminates an area that can be considerably larger than the object, resulting in very low contrast incoherently-compounded initial images. Second, the object may extend well beyond the memory-effect of the sample. Acousto-optical modulation allows to noninvasively isolate small portions of the object that can be reconstructed with high fidelity with our approach when a sufficient memory-effect is present, as we demonstrate below.

For each of the $m=1..M$ ultrasound focus positions, the $N$ measured fields (Fig.\ref{fig:acousto_optics}C) are fed as the input to the computational image-guided wavefront-shaping algorithm. Applying the same algorithm yields high contrast diffraction-limited images (Fig.\ref{fig:acousto_optics}D). 
Mosaicking the optimized images results in a wide FoV reconstructed image of the target (Fig.\ref{fig:acousto_optics}E).

\section*{Discussion}\label{discussion_sec}

We have demonstrated a holography-based computational method for noninvasive guide-star free high-resolution imaging through complex media. 
Our approach eliminates the need for guide stars or SLMs, does not require knowledge or measurement of the medium's reflection-matrix, control or knowledge of the illumination, or labeling of the target.

By leveraging state-of-the-art automatic-differentiation optimization tools used for the training of neural-networks, our physical-model based method allows an efficient parallel multi-conjugate correction of a number of DoFs significantly exceeding the number of frames, captured using a simple experimental setup. 

In the presented embodiment, only the correction phase masks were optimized. However, the optimization can be extended to additional parameters, such as the distances between correction planes, object position, and experimental imperfections. We note that while the presented implementation allows automatic determination of the object axial position, it is agnostic to its lateral position, as transverse shifts have no effect over the image metric.

In our experiments we have used randomly varying illumination to generate the required number realizations. These may also be generated by natural dynamics of the object or sample at the illumination path, as long as the scattering between the object and the camera remains constant. 
The number of captured frames can be potentially reduced by a higher pixel-count acquisition, where the incoherent compounding is of different sub-apertures rather than sequential acquisitions.
 
The challenge of imaging and sensing through complex media has great practical importance in optics, but also a deep fundamental physical nature and a long history, stretching well beyond optics, through the fields of medical ultrasound, geophysics, and radar, as some examples. 
The generality of our approach lends itself to this wide variety of applications. One can easily modify the image metric or physical model to tackle various imaging tasks, domains, and modalities, from acoustics to radar.

\section*{Methods}\label{methods_sec}

\subsection*{Mathematical formulation}\label{math_bg}
Considering an object that is described by an amplitude and phase mask, whose complex field transmission is given by $O(\mathbf{r})$. 
Our goal is to reconstruct the transmission function $|O(\mathbf{r})|^2$ using $N$ 
holographic complex-valued field images recorded at the proximal facet of the complex medium, $E_{\mathrm{prox},n}(\mathbf{r}) $,  $n=1,\dots N$. Each field is obtained when a different random and unknown spatially-coherent speckle field illumination of the object, $S_n(\mathbf{r})$. 

The forward physical model connecting the desired object pattern with the field measured by the camera at each speckle illumination is obtained by propagating the object field, $O_n(\mathbf{r})=O(\mathbf{r})S_n$, through the scattering medium. 
Considering isoplanatic scattering, i.e. assuming that the object extent is within the memory-effect range \cite{feng1988correlations, freund1988memory},  the complex medium can be modeled as a thin random phase mask, placed at a distance,  $z_\mathrm{obj}$, from the object. 
The recorded field at the proximal facet of the medium is then given by propagating the field, $O_n(\mathbf{r})$, from the object plane to the medium facet by the distance, $z_\mathrm{obj}$, and multiplying this field by the phase mask representing the scattering medium. The field at the proximal facet of the complex medium is then given by:
\begin{equation}
E_{\mathrm{prox},n}(\mathbf{r}) = \mathcal{P}_{z_\mathrm{obj}}\left\{O_n(\mathbf{r})\right\} \cdot e^{ i\phi(\mathbf{r})} \label{eq2}
\end{equation}
where  $\mathcal{P}_z(\cdot)$ represents a propagation over a distance $z$ \cite{goodman2005introduction}.

Given the holographically-measured proximal fields, $E_{\mathrm{prox},n}(\mathbf{r}) $, the iterative computational inversion process aims at simultaneously finding both the phase mask representing the complex medium, $\phi(\textbf{r})$, and the object pattern. 
This is performed by computationally emulating an image-guided wavefront-shaping experiment, where the spatial phases of a 'virtual SLM' correction phase mask are optimized to enhance the quality of the reconstructed incoherently-compounded object image.
At each iteration, $t$, the incoherently-compounded object image is reconstructed by tracing back each of the measured fields through the correction mask,  $e^{-i\phi_{t}(\textbf{r})}$, and propagating a distance $-z_\mathrm{prop}$ to the object estimated plane by the angular spectrum method \cite{goodman2005introduction}. 
Mathematically, the $n^\mathrm{th}$ reconstructed object field at the $t^\mathrm{th}$ iteration is described by:
\begin{align}
\hat{O}_{n,t}(\mathbf{r}) 
& =\mathcal{P}_{-z_\mathrm{prop}}\!\left\{E_{\mathrm{prox},n}(\mathbf{r})\cdot\exp\left(-i\phi_t(\mathbf{r})\right)\right\}\nonumber\\
 & =\mathcal{P}_{-z_\mathrm{prop}}\!\left\{ E_{\mathrm{dist},n}(\mathbf{r}) \cdot \exp\left(i\phi(\mathbf{r})-i\phi_t(\mathbf{r})\right)\right\}
\end{align}
The incoherently compounded image is formed by averaging the intensity patterns of the back-propagated fields:
\begin{equation}
    I_{t}(\mathbf{r}) = \left\langle \left| \hat{O}_{n,t}(\mathbf{r}) \right|^2\right\rangle_n \label{eq4}
\end{equation}

The computational inversion process begins by drawing an initial virtual SLM phase pattern $\phi_{0}(\textbf{r})$, with the subscript index denoting the iteration number.
To achieve a reconstruction of $|O(\mathbf{r})|^2$, the phase $\phi_t(\mathbf{r})$ is varied via a gradient-descent algorithm aimed at optimizing the image quality metric $\mathcal{Q}_t$. A large variety of image quality metrics can be used \cite{muller1974real, yeminy2021guidestar}. In the examples given in this work, we have adapted the well-established image variance metric (representing the contrast of the reconstructed image), to enable metric minimization instead of maximization as is most commonly done (see Methods section \nameref{methods_algo} for further information regarding this matter).

Once the image quality of $I_t(\mathbf{r})$ is calculated, it can be easily differentiated with respect to $\phi_t(\mathbf{r})$ using automatic differentiation. The gradient-descent algorithm works by updating the phase mask by taking a step of size $\alpha$ in the direction of optimal image quality:
\begin{equation}
\phi_{t+1}(\textbf{r})\leftarrow \phi_t(\textbf{r})-\alpha\frac{\partial\mathcal{Q}_t}{\partial\phi_t(\textbf{r})} \label{eq7}
\end{equation}

Using this updated phase, the process is iteratively repeated until a predefined maximum number of iterations is passed or if another heuristically pre-determined stopping condition is reached,  $\Delta\phi_t = \alpha\frac{\partial\mathcal{Q}_t}{\partial\phi_t(\textbf{r})}\leq  \Delta\phi_\mathrm{stop}$.

\subsection*{Experimental setup}\label{exp_setup}
A detailed description of the experimental setup 
is given in Supplementary Section 1. 
Briefly, the random realizations of illuminations were generated by illuminating an optical diffuser (Newport light shaping diffusers of either $30^\circ$ or $60^\circ$scattering angle) placed on a controllable rotation mount (Thorlabs K10CR1). In all experiments except the acousto-optic experiment, the laser source was a 20-mW Helium-Neon laser (Thorlabs HNL210L), and the rotating diffuser was placed at a small distance ($\sim 1.4$cm) from the USAF-1951 resolution target (Thorlabs R1DS1N). In the acousto-optic experiment (Fig.\ref{fig:acousto_optics}), the laser used was a $532$nm pulsed laser (Standa Ltd., Vilnius, Lithuania). 

In the experiment presented in Fig.\ref{fig:optim_evolution}, a diffuser ($60^\circ$ scattering angle, Newport) was used as the scattering layer, positioned 1.5cm in front of the target. In  Fig.\ref{fig:2 diffusers}, two optical diffusers ($5^\circ$ scattering angle, Newport), separated by $\sim300\mu$m, were used as the scattering layer, positioned 3.1cm from the target.
In Fig.\ref{fig:fiber_results}B-E, the complex medium was a Fujikura MCF bundle (FIGH-50-1100N) with $50$k cores and $1.1$mm diameter. In Fig.\ref{fig:fiber_results}F-I, the complex medium was a Schott MCF bundle (1533385) with $18$k cores, pixel size of $8\mu$m, and $1.1$mm diameter.
Both MCF bundles were positioned 5mm from the target. Results at other distances are presented in Supplementary Figure 3. 

In all experiments except for the acousto-optic experiment, a 4-f telescope composed of a microscope objective (Olympus PLN 20X - 1-U2B225), and an $f=100$mm tube lens, images the proximal facet of the complex medium on the camera plane (Andor Zyla 5.5 sCMOS).
The results presented in Fig.\ref{fig:acousto_optics} were taken using an experimental setup described in detail in  \cite{rosenfeld2021acousto}. Briefly, the rotating diffuser used for generating the speckle illumination was composed of two stacked $1^\circ$ optical diffusers, followed by a static $5^\circ$ optical diffuser. The target was positioned $5$cm from the static diffuser and $15.5$cm from another $5^\circ$ diffuser.

\subsection*{Reconstruction algorithm} \label{methods_algo}
For the implementation of the gradient-descent algorithm, the PyTorch library was used \cite{paszke2017automatic}, in order to leverage its automatic-differentiation and GPU-enabled capabilities. The algorithm used to perform the optimization and generate the results was a standard manually-implemented gradient-descent algorithm tailored for the specific task, as described in Supplementary Section 6.

The initial phase mask was a flat phase mask, i.e. a matrix of zeroes.
The step size taken at each iteration was an adaptive Two-Point step size as outlined in \cite{barzilai1988two}. The stopping condition, $\Delta\phi_\mathrm{stop}$, was set heuristically per complex medium (e.g., for the ordered MCF $\Delta\phi_\mathrm{stop}\approx3.1$mrad was chosen).

The image quality metric used to generate the results presented in the figures was chosen as the image variance given by: $\sigma^2_t=\frac{1}{M-1}\sum_\textbf{r}(I_t(\textbf{r}) - \bar{I_t})^2$, where $I_t(\textbf{r})$ is the reconstructed image at the $t^\mathrm{th}$ iteration and $M$ is the total number of pixels in the image. To work with converging rather then diverging values, the image variance was "wrapped" in another function for the optimization: $\mathcal{Q}_t=\tanh \left(\frac{\sigma^2_0}{\sigma^2_t} \right) / \tanh(1)$, where $\tanh(\cdot)$ is the hyperbolic tangent function. As can be immediately noticed, at $t=0$, the value is 1. This function can easily support any other valid image quality metric and only serves for convenience. Other valid image quality metrics were tested, such as image entropy and power spectrum variance, yielding similar results.

The measured running time of the results presented in Fig.\ref{fig:optim_evolution}, running on a consumer NVIDIA GeForce RTX 3090 GPU, was 6 minutes for 500 iterations. Down-sampling the same results from $600\times600$ DoF to $200\times200$ DoF, and running for the same number of iterations leads to a very good reconstruction after 10 seconds (see Supplementary Section 8). 
We note, however, that a considerably lower number of iterations yield virtually unnoticeable differences in the results, as can be seen in Fig.\ref{fig:optim_evolution}D. In fact, setting $\Delta\phi_\mathrm{stop}\approx2\textrm{mrad}$ and the image quality metric to power spectrum variance, yields unnoticeable results for the same down-sampled version, but takes only $\sim 7$ seconds and stops after 241 iterations (see Supplementary Section 8).
\backmatter

\bmhead{Supplementary information}
The supplementary information document contains details of the experiments and the source code of the presented algorithm.

\bmhead{Acknowledgments}
We acknowledge G. Weinberg and Y. Slobodkin for fruitful discussions.
This work has received funding from the European Research Council under the European Union’s Horizon 2020 
Research and Innovation Program grant number 101002406. 
\newpage

\section*{Declarations}

\bmhead{Authors' contributions}
O.K. conceived the idea. O.K., O.H., and J.B.L. designed the experimental setup. O.H. and J.B.L. performed numerical simulations and data analysis. O.H. designed and implemented the automatic-differentiation gradient-descent algorithm. J.B.L. led the experimental work. J.B.L and O.H. performed the experiments and analyzed the data under the supervision of O.K.. All authors wrote the manuscript.

\bmhead{Competing interests}
The authors declare no competing interests.

\bigskip

\bibliography{bibliography}

\end{document}